\documentclass[a4paper,twoside,10pt]{letter}
\usepackage{graphicx,saj,multicol,subeqnarray}
\def\p0{\phantom{0}}
\newcommand{\arcsec}{\ensuremath{^{\prime\prime}}}
\newcommand{\citep}{\textbf{?}}

\def\papertype{Original Scientific Paper}

\def\udc{...}
\begin{document}
\baselineskip=3.1truemm
\columnsep=.5truecm
\newenvironment{lefteqnarray}{\arraycolsep=0pt\begin{eqnarray}}
{\end{eqnarray}\protect\aftergroup\ignorespaces}
\newenvironment{lefteqnarray*}{\arraycolsep=0pt\begin{eqnarray*}}
{\end{eqnarray*}\protect\aftergroup\ignorespaces}
\newenvironment{leftsubeqnarray}{\arraycolsep=0pt\begin{subeqnarray}}
{\end{subeqnarray}\protect\aftergroup\ignorespaces}
%

% Running titles

\markboth{NEW 6 AND 3-CM RADIO-CONTINUUM MAPS OF THE SMC. PART I -- THE MAPS}
{\eightrm E.~J.~Crawford, M.~D.~Filipovi\'c, A.~Y.~De~Horta, G.~F. Wong, N.~F.~H. Tothill, D.~Dra\v skovi\'c, J.~D. Collier, T.~J. Galvin}

{\ }

\publ

\type

{\ }

% Title

\title{New 6 and 3-cm Radio-continuum maps of the Small Magellanic Cloud. Part I -- The Maps}

% Authors

\authors{E.~J.~Crawford, M.~D.~Filipovi\'c, A.~Y.~De~Horta, G.~F.Wong,}
\authors{N.~F.~H. Tothill, D.~Dra\v skovi\'c, J.~D. Collier, T.~J.~Galvin}

\vskip3mm

% Address

\address{University of Western Sydney, Locked Bag 1797, Penrith South DC, NSW 2751, AUSTRALIA}

\Email{e.crawford}{uws.edu.au}

%\address{$^2$Institute for Astrophysics, Department of Physics and Astronomy,
%University of New Mexico\break 800 Yale Blvd., N.E., Albuquerque, NM
%87131 USA}

%\Email{neb}{duric.org}

% Received and Accepted dates

\dates{October XX, 2011}{October XX, 2011}

% Abstract

\summary{We present new 6 and 3-cm radio-continuum maps of the Small Magellanic Cloud (SMC), created with the ``peeling'' technique and a joint deconvolution. The maps have resolutions of 30\arcsec{} and 20\arcsec{} and r.m.s. noise of 0.7 and 0.8~mJy/beam at 6 and 3~cm, respectively. These maps will be used for future studies of the SMC's radio source population and overall extended structure.}

% Keywords (see keywords.pdf file)


\begin{multicols*}{2}
{

% Sections

\section{1. INTRODUCTION}
 \label{s:intro}

In the past two decades, several Parkes, Molonglo Observatory Synthesis Telescope (MOST) and Australia Telescope Compact Array (ATCA) moderate resolution ($\sim$arcmin) surveys of the Magellanic Clouds (MCs) have been completed (Filipovi\'c et al. 1998).

Deep ATCA and Parkes radio-continuum and snap-shot surveys of the Small Magellanic Cloud (SMC) were conducted at 20, 13, 6 and 3~cm by Filipovi\'c et al (2002), achieving sensitivities of 1.8, 0.4, 0.8 and 0.4~mJy/beam respectively. These surveys were conducted in mosaic mode with between 35 and 320 separate pointings using 5 antennae in the 375-m array configuration, with short-spacing information from single-dish Parkes observations. The maps have angular resolutions of 98\arcsec, 40\arcsec, 30\arcsec\ and 20\arcsec\ at the respective wavelengths listed above. Recently, we published a set of new high resolution (few-arcsec) radio-continuum mosaic images of the SMC at 20~cm created by combining observations from ATCA and Parkes (Wong et al. 2011). The latest images of the SMC at 6 and 3~cm were presented by Dickel et al. (2005, 2010). These maps display significant artefacts due to the way they were constructed as linear mosaics of individually cleaned images. By jointly deconvolving these data with other archival data we were able to obtain significantly improved maps of the SMC.

\section{2. IMAGING TECHNIQUES}
 \label{s:imag}

When the area to be imaged exceeds the field of view of the telescope, a mosaic must be made. There are two methods for assembling a mosaic: Image each field and combine the images at the end; or combine the data at the beginning and image them jointly. The field-by-field approach has the advantage that it has modest processing requirements. However, not all available information in the data is utilised. The downside to the joint approach is a huge increase in the processing time and complexity.

A further problem with wide field imaging is the effect of ``off-axis'' sources. A source that is outside the primary beam of the telescope can still influence the response of the receiver. The solution to this problem is to use the  ``peeling'' technique detailed in Hughes et al. (2007). Essentially, the techniques of self-calibration and source subtraction are used to correct for off-axis sources and thus reduce the artefact levels.

The interferometer images only sample the Fourier plane down to the shortest baseline --- the largest scale structures are missed. To correct for this deficiency, archival ``zero-spacing'' data from Parkes (a large single dish) observations (Filipovi\'c et al. 1997) have been incorporated into all of the new survey images. However, these observations do not cover the whole field of the ATCA observations. The Parkes data also have much higher r.m.s. noise than the ATCA data. 

\vspace{1em}
\centerline{{\bf Table 1.} ATCA Projects used.}
\vskip2mm
\centerline{\begin{tabular}{cl}
\hline
Project Code & Dates Observed \\
\hline
C1604        & Jun 2007 \\
C1207        & February 2005 \& 2006 \\
C882\p0      & Jun 2000 \\
C859\p0      & March 2007 \\
C634\p0      & October 1997 \\
\hline
\end{tabular}}

%\vspace{1em}

\section{3. DATA REDUCTION TECHNIQUE AND RESULTS}
 \label{s:data}

The majority of archival data used come from ATCA project C1207 (Dickel et al. 2010), available in calibrated form\footnote{\texttt{http://www.atnf.csiro.au/research/smc\_ctm/}}. A search of the Australia Telescope Online Archive\footnote{\texttt{http://atoa.atnf.csiro.au}} yielded four more projects that cover the SMC at 6 and 3~cm (Table~1). All data were inspected, peeled if necessary, and then combined into a single ``dirty map''. The SDI clean algorithm (Sault \& Killeen 2010) was used to deconvolve the images, which were then restored. The new maps are presented with and without zero-spacing data in Figs.~1. to 4. The resolutions are 30\arcsec\ and 20\arcsec\ and the sensitivities of the ATCA data are 0.7 and 0.8 mJy/beam at 6~cm (Fig.~1) and 3~cm (Fig.~3), respectively. The r.m.s. noise levels in Figs.~2 and 4 are dominated by the additional noise of the single dish data (Filipovi\'c et al. 1997).  

As only project C1207 was designed as a survey of the entire region, the maps suffer from nonuniform sensitivity, a consequence of differing integration times at each pointing. Figures~5 \& 6 demonstrate how the sensitivity changes across the field.

All these images can be downloaded from  \texttt{http://spacescience.uws.edu.au/mc/smc/}

% Acknowledgements

\acknowledgements{The Australia Telescope Compact Array and Parkes radio telescope  are parts of the Australia Telescope National Facility which is funded by the Commonwealth of Australia for operation as a National Facility managed by CSIRO. This paper includes archived data obtained through the Australia Telescope Online Archive (http://atoa.atnf.csiro.au). We used the {\sc karma} and {\sc miriad} software package developed by the ATNF.}

\vspace{1em}
% References

\references

Dickel, J.R., McIntyre, V.J., Gruendl, R.A., Milne, D.K.: 2005, \journal{Astron. J.} \vol{129}, 790.

Dickel, J.R., Gruendl, R.A., McIntyre, V.J., Amy, S.W.\ 2010, \journal{Astron. J.}, \vol{140}, 1511

Filipovi\'c, M.D., Jones, P.A., White, G.L., Haynes, R.F., Klein, U., Wielebinski, R: 1997, \journal{Astron. Astrophys. Suppl. Ser.} \vol{121}, 321.

Filipovi\'c, M.D., Haynes, R.F., White, G.L., Jones, P.A.: 1998, \journal{Astron. Astrophys. Suppl. Ser.} \vol{130}, 421.

Filipovi\'c, M.D., Bohlsen, T., Reid, W., Staveley-Smith, L., Jones, P.A., Nohejl, K., Goldstein, G.: 2002, \journal{Mon. Not. R. Astron. Soc.} \vol{335}, 1085.

Hughes, A., Staveley-Smith, L., Kim, S., Wolleben, M., Filipovi\'c, M.D.: 2008, \journal{Mon. Not. R. Astron. Soc.} \vol{382}, 543.

Sault, R., Killeen, N.: 2010, Miriad Users Guide, ATNF

Wong, G.F., Filipovi{\'c}, M.D., Crawford, E.J., de Horta, A.Y., Galvin, T., Dra\v skovi{\'c}, D., Payne, J.L.: 2011, \journal{Serb. Astron. J.}, \vol{182}, 43.

\endreferences% Figures in ps or eps format with resolution 600dpi

}

\end{multicols*}
\vfill\eject

\centerline{\includegraphics[angle=-90,width=\textwidth]{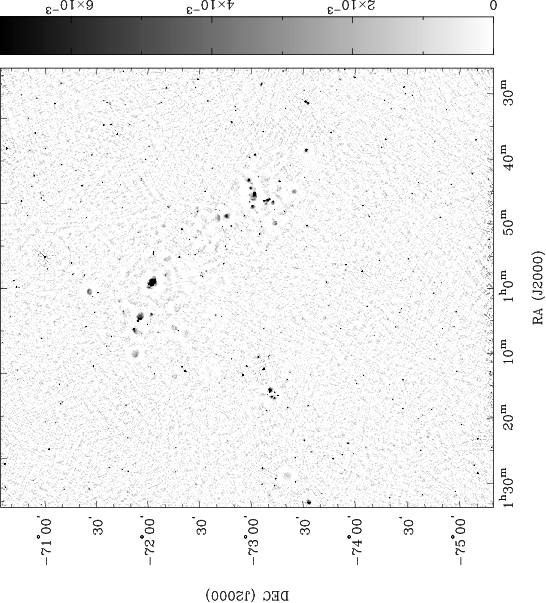}}
\figurecaption{1.}{The SMC at 6~cm without zero-spacing data. The map has a resolution of 30\arcsec. The sidebar gives the intensity scale in units of Jy/beam.}

\centerline{\includegraphics[angle=-90,width=\textwidth]{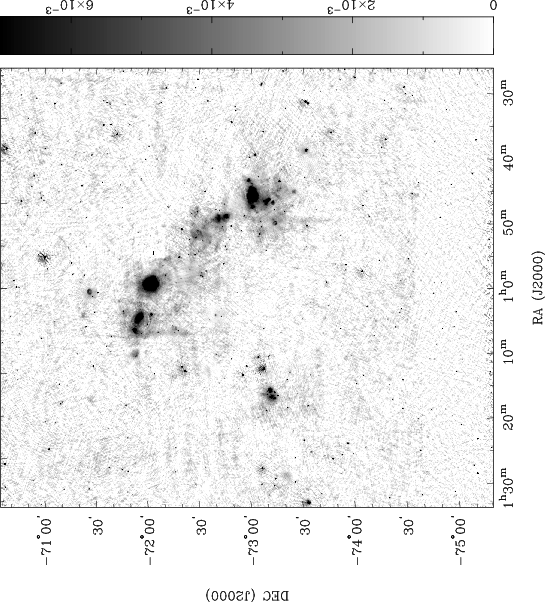}}
\figurecaption{2.}{The SMC at 6~cm with zero-spacing data. The map has a resolution of 30\arcsec. The sidebar gives the intensity scale in units of Jy/beam. The zero-spacing data does not cover the same spatial area as the interferometer data.}

\centerline{\includegraphics[angle=-90,width=\textwidth]{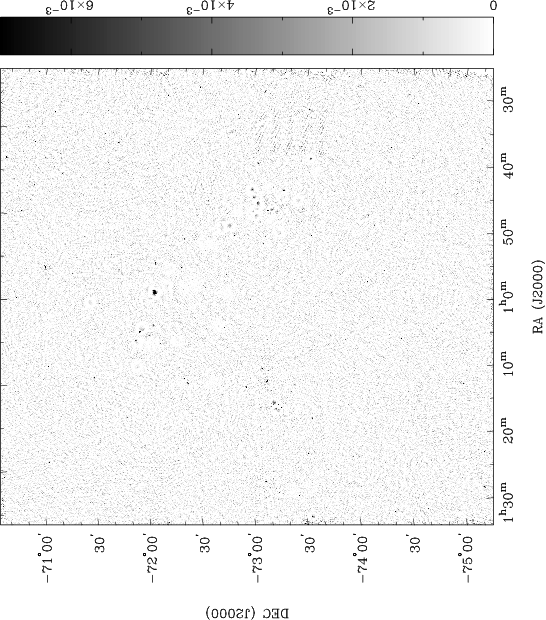}}
\figurecaption{3.}{The SMC at 3~cm without zero-spacing data. The map has a resolution of 20\arcsec. The sidebar gives the intensity scale in units of Jy/beam. }

\centerline{\includegraphics[angle=-90,width=\textwidth]{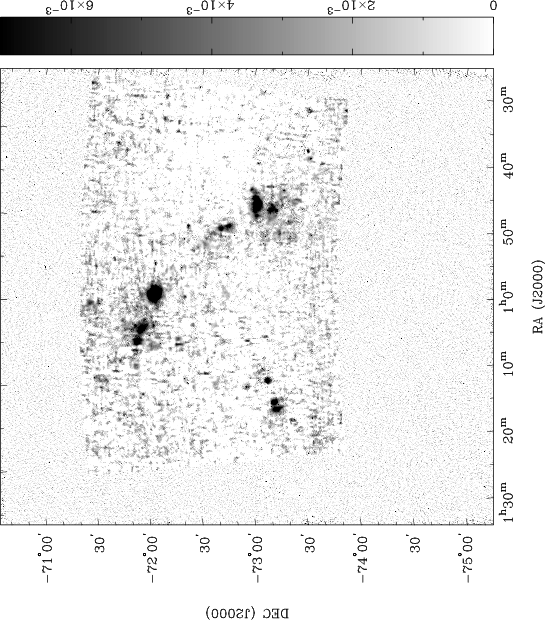}}
\figurecaption{4.}{The SMC at 3~cm with zero-spacing data. The map has a resolution of 20\arcsec. The sidebar gives the intensity scale in units of Jy/beam. The zero-spacing data does not cover the same spatial area as the interferometer data.}

\centerline{\includegraphics[angle=-90,width=\textwidth]{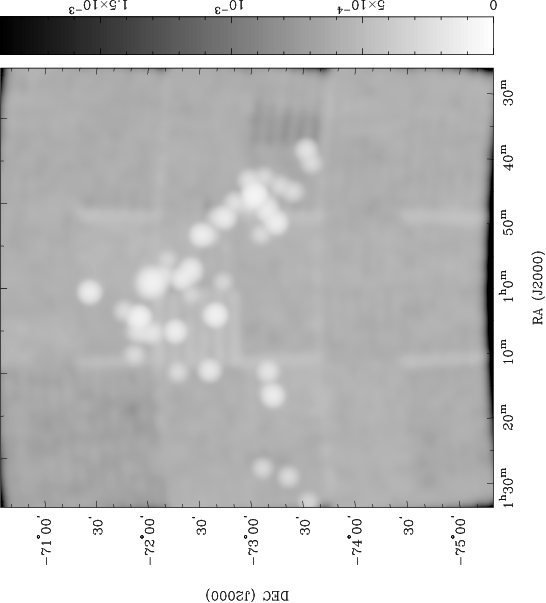}}
\figurecaption{5.}{Theoretical r.m.s. sensitivity map of the 6~cm data used in this study in Jy/beam. Lighter areas indicate higher sensitivity.}

\centerline{\includegraphics[angle=-90,width=\textwidth]{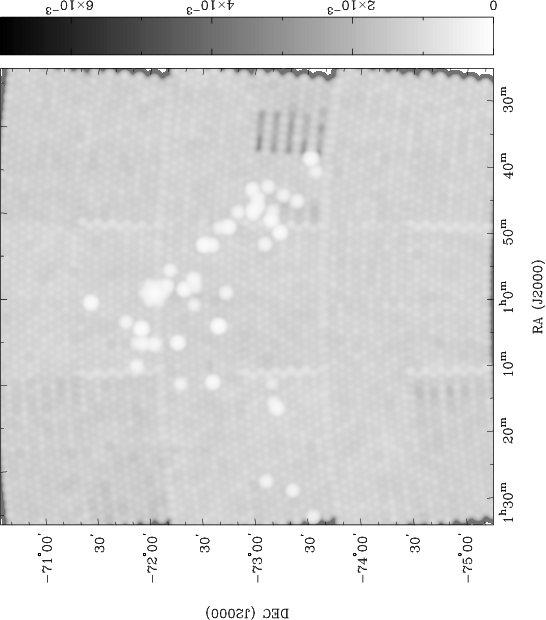}}
\figurecaption{6.}{Theoretical r.m.s. sensitivity map of the 3~cm data used in this study in Jy/beam. Lighter areas indicate higher sensitivity.}

\vfill\eject

{\ }

% Serbian abstract

% Title

\naslov{NOVO PROUQAVA{NJ}E MALOG MAGELANOVOG OBLAKA U RADIO-KONTINUMU NA 6~CM i 3~CM: DEO~{\bf I} - MAPE}

% Authors

\authors{E.~J.~Crawford, M.~D.~Filipovi\'c, A.~Y.~De~Horta, G.~F.Wong,}
\authors{N.~F.~H. Tothill, D.~Dra\v skovi\'c, J.~D. Collier, T.~J.~Galvin}

\vskip3mm

% Address

\address{University of Western Sydney, Locked Bag 1797, Penrith South DC, NSW 2751, AUSTRALIA}

\vskip.7cm

% UDC

\centerline{UDK \udc}

% Papertype

\centerline{\rit }

\vskip.7cm

\begin{multicols}{2}
{

% Abstract

\rrm 

U ovoj studiji predstav{lj}amo nove {\rm ATCA} rezultate posmatra{nj}a visoke rezolucije i oset{lj}ivosti u radio-kontinumu Malog Magelanovog Oblaka (MMO) na {\rm $\lambda$=6~cm ($\nu$=4.8~GHz) i $\lambda$=3~cm ($\nu$=8.64~GHz)}. Nove radio-mape nastale su spaja{nj}em arhivskih mozaik posmatra{nj}a tzv. {\rm ``peeling''} tehnikom i to na 6~cm i 3~cm. Naxe nove mape imaju rezoluciju od $\sim$30\arcsec\ (6~cm) i $\sim$20\arcsec\ (3~cm) i oset{lj}ivost od {\rm r.m.s.=0.7~mJy/beam}. Ove mape {\cc}e biti korix{\cc}eni u budu{\cc}im istra{\zz}iva{nj}ima kako objekata tako i ukupne strukture \mbox{MMO-a}.

}
\end{multicols}

\end{document}